\documentclass[epj]{webofc}
\usepackage[utf8]{inputenc}
\usepackage[varg]{txfonts}   % Web of Conferences font
\usepackage{booktabs}
\usepackage{xcolor}
\usepackage{subcaption}

\definecolor{darkred}{rgb}{0.4,0.0,0.0}
\definecolor{darkgreen}{rgb}{0.0,0.4,0.0}
\definecolor{darkblue}{rgb}{0.0,0.0,0.4}
\usepackage[bookmarks,linktocpage,colorlinks,
    linkcolor = darkred,
    urlcolor  = darkblue,
    citecolor = darkgreen]{hyperref}
\wocname{EPJ Web of Conferences}
\woctitle{Lattice2017}
\providecommand*{\Psibar}{\ensuremath{\overline{\Psi}}}
\providecommand*{\ord}{\ensuremath{\text{O}}}
\providecommand*{\I}{\ensuremath{\mathbf{1}}}
\providecommand*{\e}{\ensuremath{\text{e}}}
\providecommand*{\de}{\ensuremath{\text{d}}}

\newcommand{\quotes}[1]{``#1''}

\begin{document}
%%%%%%%%%%%%%%%%%%%%%%%%%%%%%%%%%%%%%%%%%%%%%%%%%%%%%%%%%%%%%%%%%%%%%%%%%%%%%
%
\selectlanguage{english}
\vspace{-2cm}
\begin{flushright}
CP3-Origins-2017-041
\end{flushright}
\vspace{-.2cm}
%
%----------------------------------------------------------------------------
\title{%
Electromagnetic corrections to the hadronic vacuum polarization of
the photon within QED$_{\rm L}$ and QED$_{\rm M}$
}
%----------------------------------------------------------------------------
\author{%
\firstname{Andrea} \lastname{Bussone}
\inst{1}
\fnsep 
\thanks{Speaker, \email{bussone@cp3-origins.net}}
\and
\firstname{Michele} \lastname{Della Morte}\inst{1} \and
\firstname{Tadeusz}  \lastname{Janowski}\inst{1}
% etc.
}
%----------------------------------------------------------------------------
\institute{%
CP$^3$-Origins, University of Southern Denmark, Campusvej 55, DK-5230 Odense M, Denmark
}
%----------------------------------------------------------------------------
\abstract{%
  We compute the leading QED corrections to the hadronic vacuum polarization (HVP) of the photon, relevant for the determination of leptonic anomalous magnetic moments, $a_\ell$. We work in the electroquenched approximation and use dynamical QCD configurations generated by the CLS initiative with two degenerate flavors of non-perturbatively O($a$)-improved Wilson fermions. We consider QED$_{\rm L}$ and QED$_{\rm M}$ to deal with the finite-volume zero modes. We compare results for the Wilson loops with exact analytical determinations. In addition we make sure that the volumes and photon masses used in QED$_{\rm M}$ are such that the correct dispersion relation is reproduced by the energy levels extracted from the charged pions two-point functions. Finally we compare results for pion masses and the HVP between QED$_{\rm L}$ and QED$_{\rm M}$. For the vacuum polarization, corrections with respect to the pure QCD case, at fixed pion masses, turn out to be at the percent level.
}
%----------------------------------------------------------------------------
\maketitle
%----------------------------------------------------------------------------
%
\vspace{-0.3cm}
\section{Introduction}\label{introduction}
The $(g-2)_\mu$ is one of the most precise measurement in particle physics and it serves as a stringent test of the Standard Model (SM).
The persistent $3-4$ $\sigma$ discrepancy between the experimental value and theoretical calculation \cite{Amsler:2008zzb} has generated a lot of interests in the past years.
The dominant contribution to the anomalous magnetic moment is due to  QED and at the level we are today in the experiment we need to include 
all the possible contributions from the SM.
%
%Recently, $Z'$ models were introduced to explain the violations of lepton flavor universality because of tensions in the measurements of 
%$R_K$ and $R_{K^\star}$.
%For example in Ref.~\cite{DiChiara:2017cjq} is pointed out that a vectorial
%coupling to the $Z'$ could also alleviate the discrepancy between the $(g-2)_\mu$ measurement and the SM prediction.\\
In addition, a number of SM extensions have been recently proposed addressing 
the violations of lepton flavor universality observed in the measurements of $R_K$ and $R_{K^\star}$.
For example in Refs.~\cite{DiChiara:2017cjq, DAmico:2017mtc} is pointed out that in models with a vectorial coupling to a $Z'$ or 
in some fundamental composite Higgs model the 
discrepancy between the $(g - 2)_\mu$ measurement and the SM prediction can be alleviated,
while explaining at the same time the flavor anomalies.
The lattice regularization can provide a non-perturbative  determination of the
hadronic vacuum polarization contribution to the muon magnetic anomaly, that is the one dominating the error and represents the second most important contribution.
Especially in view of the
new planned experiments E989 at FNAL and E34 at J-PARC that will improve the determination of $a_\mu$ by a factor four on the experimental side.
The dispersive approach to calculate the leading hadronic contribution to the muon anomaly is still the most accurate, and it obviously contains all the SM contributions.
In order to make contact with it we need to take into account QED ef\mbox{}fects.
As a further motivation, in Ref.~\cite{Calame:2015fva} an alternative method to measure the hadronic
contributions using experimental data employing a space-like kinematics is proposed, which allows
for a direct comparison with lattice estimates.\\
Finally we also point out that the Light-by-Light (LbL) contribution to the anomaly, $a_\mu^{\rm LbL}$,
is of the same order in the electromagnetic coupling $\alpha$ expansion as the hadronic leading order one with an extra insertion 
of a photon line, so the two terms must be considered at the same time.

In this work we present preliminary results on the electromagnetic corrections to the HVP.
The $3\sigma$ discrepancy mentioned above
translates in a 4\% ef\mbox{}fect on $a_\mu^{\rm HLO}$, furthermore, 
the lattice estimates of such a pure QCD term have currently uncertainties around 5\%.
It is then natural to ask whether one can isolate a correction of order 1\%, which is the 
expected size for a QED contribution. That is the main issue we try to address here, since
such a correction can in fact almost completely resolve the theory vs experiment discrepancy.
A recent work in the direction of including QED ef\mbox{}fects, as well
as strong isospin breaking corrections,
has been presented in Ref.~\cite{Boyle:2017gzv}.

In Section~\ref{QED_lattice} we discuss our choices of finite-volume zero modes regularization. There we present a test of the quenched QED (qQED) configurations comparing results for Wilson loops with the infinite volume analytical calculations.
After that we present preliminary results on the Q(C+qE)D pseudoscalar sector and further details on the QED$_{\rm M}$ in order to make sure that photon masses allow to reproduce the correct dispersion relations.
In Section~\ref{em_corr_muon_anomaly} we present our results for the EM corrections to the muon anomaly. We present a new emerging strategy that gives us a direct access to the EM correction on the scalar Vacuum Polarization and allows us to determine EM contributions to the anomaly without large systematic ef\mbox{}fects.
\vspace{-0.3cm}
\section{QED on the lattice}\label{QED_lattice}
We do not discuss here all the issues with QED on the lattice, for a review on that the reader is referred to Refs.~\cite{Patella:2017fgk, Portelli:2013jla}.
Let us restrict to the case of the non-compact formulation, since it will be the one considered in the rest of the work.
QED on the lattice is plagued by the well-known zero-mode problem.
As a consequence of it in a periodic lattice charged states are forbidden to propagate \cite{Hayakawa:2008an}.\\
In order to correct for that we choose to employ
QED$_{\rm L}$ \cite{Borsanyi:2014jba} and QED$_{\rm M}$ \cite{Endres:2015gda}
as IR-regularizations.\\
QED$_{\rm L}$ has a positive definite Hamiltonian although the conditions to remove the spatial zero-modes result in a non-local constraint\footnote{The potential issues and implications are presented in Ref.~\cite{Patella:2017fgk}.
}.
Due to the nature of QED important finite-volume ef\mbox{}fects are expected to be found, i.e.~the corrections are power-like.\\
In QED$_{\rm M}$ the spatial zero-modes are regulated, as in Perturbation Theory (PT), with a Gaussian weight, and the theory has all the advantages of a local Quantum Field Theory.
The introduction of a photon mass term solves as well the Gauss's law problem. 
Finite volume corrections in this case are expected to be exponentially suppressed with the photon mass, but an extra power-like extrapolation
to vanishing mass, $m_\gamma\rightarrow 0$, has to be performed.
\subsection{qQED Wilson loops}
\begin{figure}[!ht]
%\begin{center}
\begin{subfigure}{0.5\textwidth}
\includegraphics[scale=0.5]{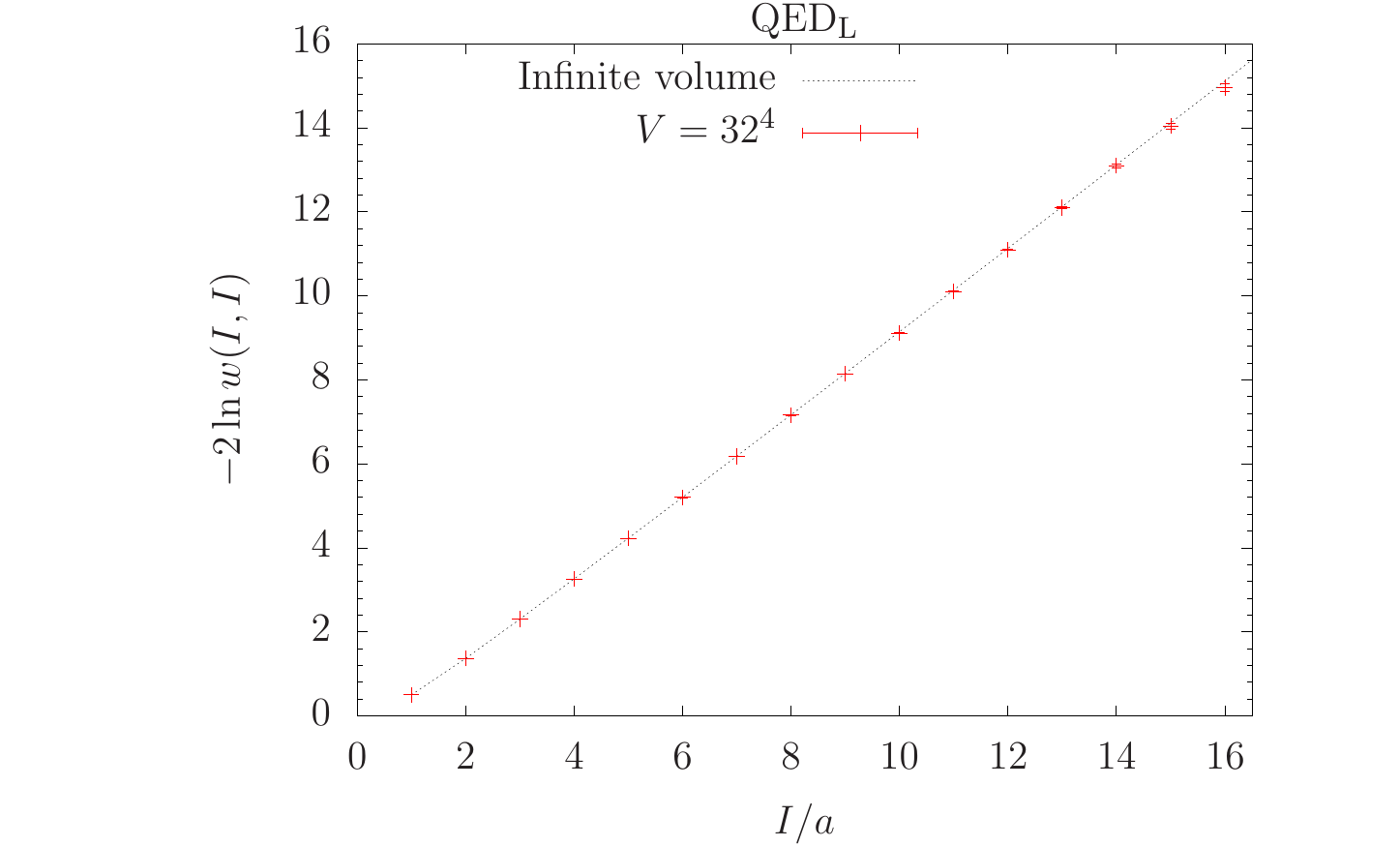} 
\captionsetup{width=.95\linewidth}
\caption{Wilson loop values in the finite lattice (red segments) and the infinite lattice predictions (dashed line).}
\label{fig:wilson_loops_qedl}
\end{subfigure}
%\hspace{-0.8cm}
\hfill
\begin{subfigure}{0.5\textwidth}
\includegraphics[scale=0.5]{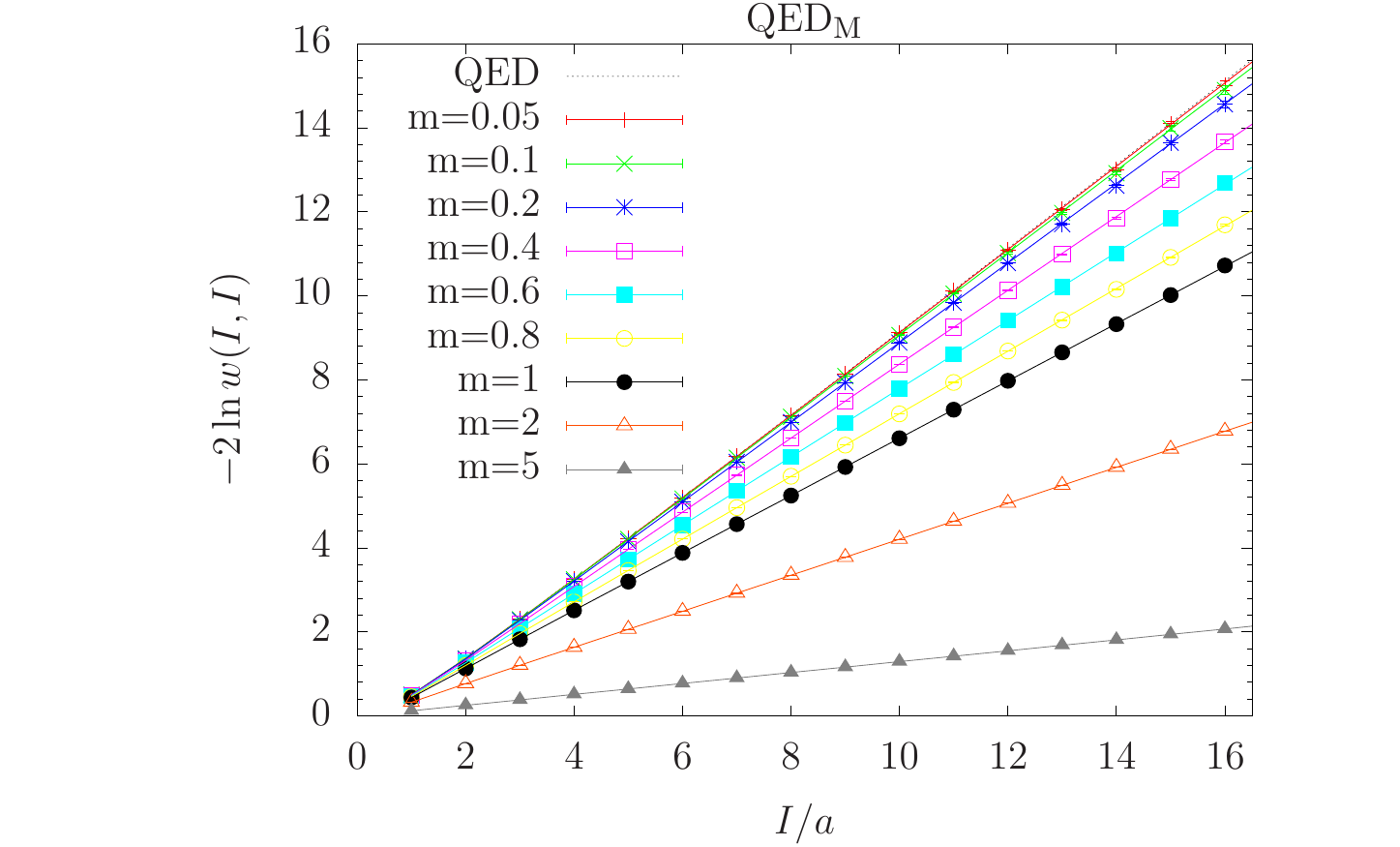}
\captionsetup{width=.95\linewidth}
\caption{Wilson loop values in the massless limit (dashed line), massive ones (solid lines) and corresponding values in the finite lattice.}
\label{fig:fig:wilson_loops_qedm}
\end{subfigure}
\caption{Comparison between the square Wilson loop values in infinite volume and the finite lattice, $V=32^4$, for the two dif\mbox{}ferent IR-regularizations. The photon mass $m$ is given in lattice units.}
\label{fig:wilson_loops}
%\end{center}
\vspace{-0.7cm}
\end{figure}

We tested the code for the generation of qQED configurations by comparing the Wilson loops expectation values 
in an infinite lattice and the one in the finite volume.
The square Wilson loops expectation values, $w_{\mu\nu}(I,I)$, with 
$I$ side length in the $(\mu,\nu)$ plane, can be exactly calculated in the infinite volume, and key ingredients are given by
\begin{align}
w_{\mu\nu} (I,I) &= \exp \left( 2e^2Q^2  \left[ C_\mu(I,0) - C_\nu(I, I\hat{\nu})  \right] \right),\\
C_\mu(I,x) &= I D(x) + \sum_{\tau=1}^{I-1} (I-\tau) D(x+\tau\hat{\mu}),
\end{align}
where $e$ is the electric charge and $Q$ the quark charge in units of $e$.
The formulae assume the knowledge of the infinite volume coordinate space quark propagator, i.e.~$D(x)$, and they can be applied in both massless and massive case.
The first is calculated in QED$_{\rm L}$ through the L\"uscher-Weisz algorithm \cite{Luscher:1995zz}, while in QED$_{\rm M}$ the Borasoy-Krebs algorithm is employed \cite{Borasoy:2005has}.\\
The result of the tests is shown in Fig.~\ref{fig:wilson_loops}, where the logarithm of the square Wilson loops, $w(I,I)$, averaged over the directions $\mu,\nu$, are plotted against the side length for both the infinite volume prediction and a $32^4$ volume.
%%%%%%%%%%%%%%%%%%%%%%%%%%%%%%%%%%%%%%%%%%%%%%%%%%%%%%%%%%%%
%%%%%%%%%%%%%%%%%%%%%%%%%%%%%%%%%%%%%%%%%%%%%%%%%%%%%%%%%%%%
%
\subsection{Q(C+qE)D pseudoscalar spectrum}
In this exploratory study on the electromagnetic corrections of the HVP we work in the electroquenched approximation and 
use dynamical QCD configurations generated by the CLS initiative with two degenerate flavors of non-perturbatively $\ord(a)$ improved Wilson fermions \cite{Fritzsch:2012wq}.
The relevant parameters for the QCD ensembles are given in Tab.~\ref{tab:qcd_ensembles}
\begin{table}[thb]
  \centering
  \caption{QCD ensemble parameters and results taken from Ref.~\cite{Capitani:2015sba}.}
  \label{tab:qcd_ensembles}% Give a unique label
  \begin{tabular}{llllllllll}\toprule
    Run & $L/a$ & $\beta$ & $c_{sw}$ & $\kappa$ & $\kappa_c$ & $am_\pi$ & $m_\pi L$ & $a$[fm] & $m_\pi$[MeV] \\\midrule
  A3 & 32 & 5.20 & 2.01715 & 0.13580 & 0.1360546 & 0.1893(6) & 6.0 & 0.079(3)(2) & 473\\
  A4 & 32 & 5.20 & 2.01715 & 0.13590 & 0.1360546 & 0.1459(6) & 4.7 & 0.079(3)(2) & 364\\
  A5 & 32 & 5.20 & 2.01715 & 0.13594 & 0.1360546 & 0.1265(8) & 4.0 & 0.079(3)(2) & 316\\\bottomrule
  \end{tabular}
% \vspace{-0.5cm}
\end{table}
 We add to preexistent QCD configurations the qQED ones by forming a $\mathbf{U}(3)$ gauge theory with un-improved
fermions.
We implemented QED$_{\rm L}$ and QED$_{\rm M}$ at the physical value of electric charge with quark charges $Q_u = 2/3$ and $Q_d= -1/3$, respectively for the up and down quark.\\
In Tab.~\ref{tab:qcd_ensembles} preliminary results on the pseudoscalar masses are given. Those are calculated by 
plateau-fitting cosh-like ef\mbox{}fective masses.
We used point-sources and the errors are calculated through a  single-elimination jackknife procedure.
\begin{table}[thb]
  \centering
  \caption{Pseudoscalar masses in Q(C+qE$_{\rm L}$)D (denoted by $m_\gamma = 0 $) and Q(C+qE$_{\rm M}$)D. 
  The resulting pion masses go from about 380 MeV to about 640 MeV.}
  \label{tab:qced_ensembles}% Give a unique label
  \begin{tabular}{lllllll}\toprule
   Run & $a m_\gamma$ & $m_\gamma L$ & $am_{\pi^0, \, u\overline{u}}$ & $am_{\pi^0, \, d\overline{d}}$ & $am_{\pi^-} = am_{\pi^+}$ & $N_{\rm cnf}$\\\midrule
  A3 & 0 & x & .2549(9) & .2071(9) & .2330(9) & 312 \\
  A3 & 0.1 & 3.2 & .2556(7) & .2074(8) & .2337(8) & 330 \\
  A3 & 0.25 &  8.0 & .2553(7) & .2072(8) & .2331(8) & 330 \\
  \hline
  A4 & 0 & x & .2240(8)  & .1691(9) & .1994(9) & 400 \\
  A4 & 0.1 &  3.2 & .2252(9) & .1699(9) & .2005(9) & 380 \\
  A4 & 0.25 &  8.0 & .2246(8) & .1700(10) & .1998(9) & 380 \\
  \hline
  A5 & 0 & x & .2105(7)  & .1526(9) & .1849(8) & 501 \\
  A5 & 0.1 &  3.2 & .2114(7) & .1528(9) & .1856(8) & 481 \\
  A5 & 0.25 &  8.0 & .2111(7) & .1531(9) & .1852(8) & 481 \\\bottomrule
  \end{tabular}
 % \vspace{-0.7cm}
\end{table}

\noindent Notice that for such chiral masses in QCD, i.e.~$m_0\simeq m_c^{\rm QCD}$, a change on the critical mass, $m_c$, of about 1\% may translate into a change of about 100\% in the quark masses\footnote{We recall that we are working with Wilson fermions.}, hence a huge change in pion masses.
Most importantly the charged pion masses in the A5 Q(C+qE)D ensemble match the A3 QCD ones.
This will be be useful later when computing the EM correction to the HVP.
Finally we remark that finite volume and photon mass ef\mbox{}fects have been checked with PT formulae and are negligible within errors.
\subsection{Further details on QED$_{\rm M}$}
\begin{figure}[!ht]
\begin{center}
\includegraphics[scale=0.6]{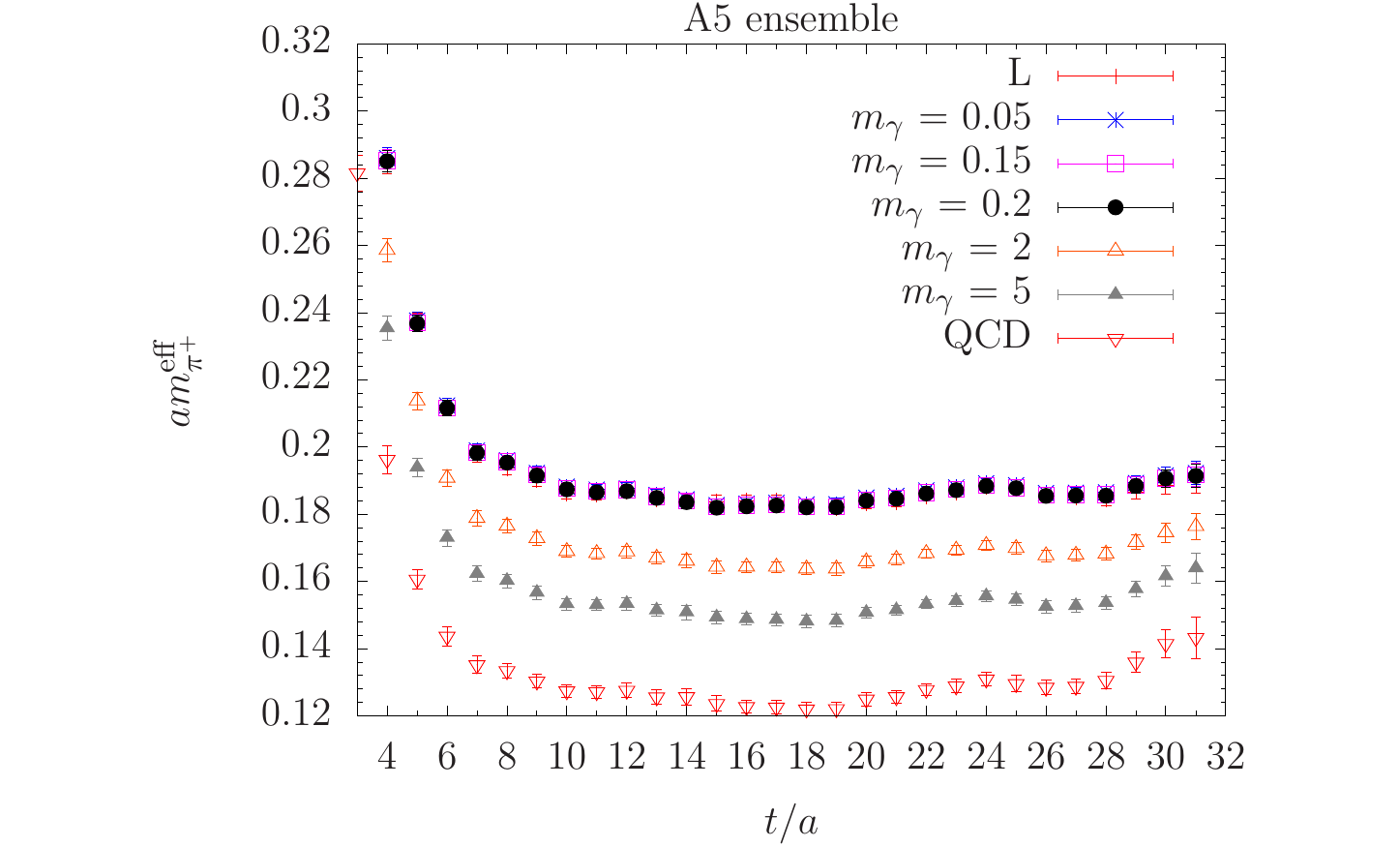}
\caption{
Ef\mbox{}fective charged pseudoscalar masses
for dif\mbox{}ferent photon masses and comparison with
QCD and Q(C+E$_{\rm L}$)D.
Notice no $t$-linear term is present for the values of  $m_\gamma$ explored.}
\label{fig:eff_masses_phot_mass}
\end{center}
\vspace{-0.7cm}
\end{figure}

\begin{figure}[!ht]
%\begin{center}
\begin{subfigure}{0.5\textwidth}
\includegraphics[scale=0.5]{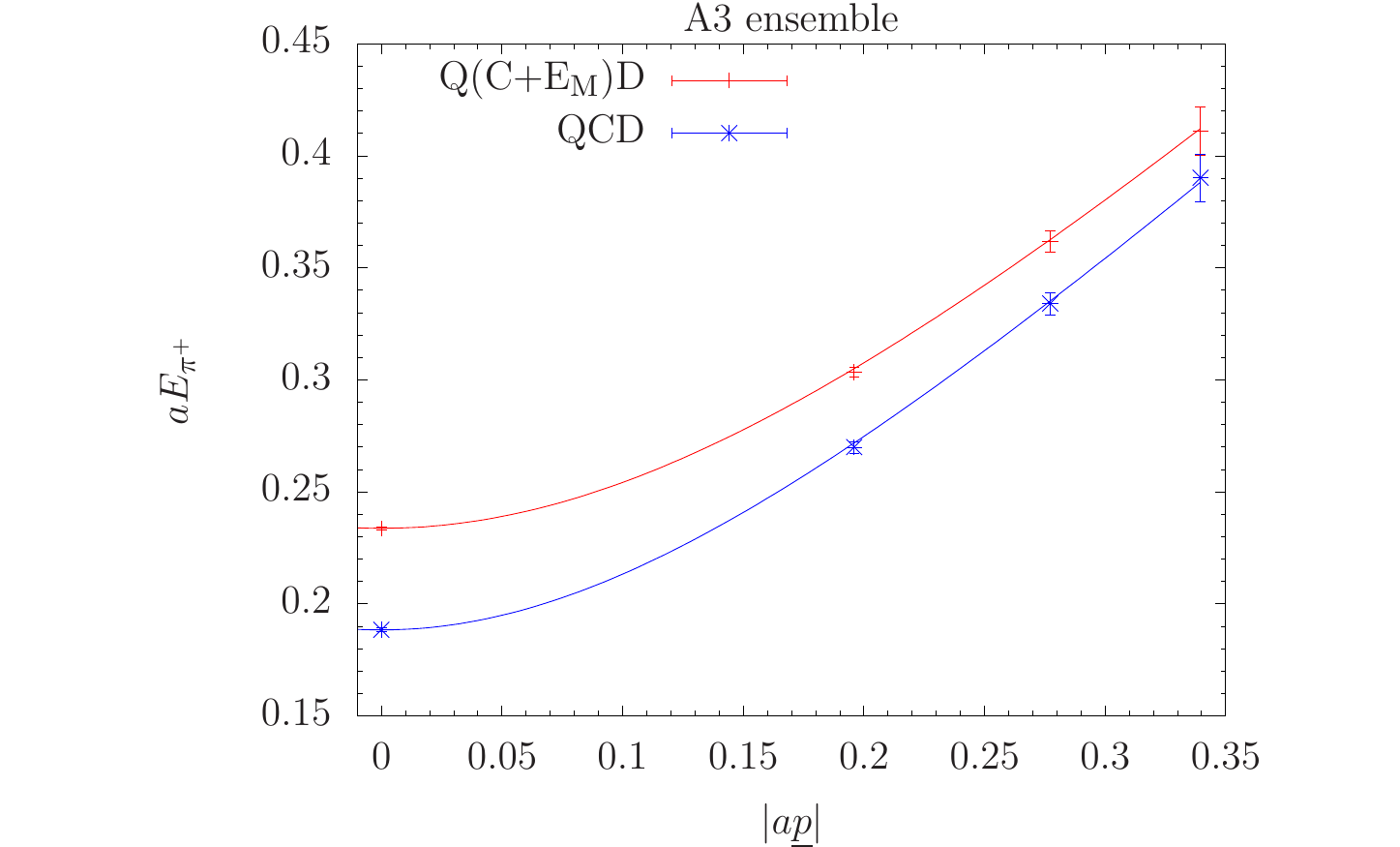} 
\captionsetup{width=.95\linewidth}
\caption{Dispersion relation for the lightest charged
pseudoscalar meson in the A3 ensemble.}
\label{fig:disp_rel_gen}
\end{subfigure}
%\hspace{-0.8cm}
\hfill
\begin{subfigure}{0.5\textwidth}
\includegraphics[scale=0.5]{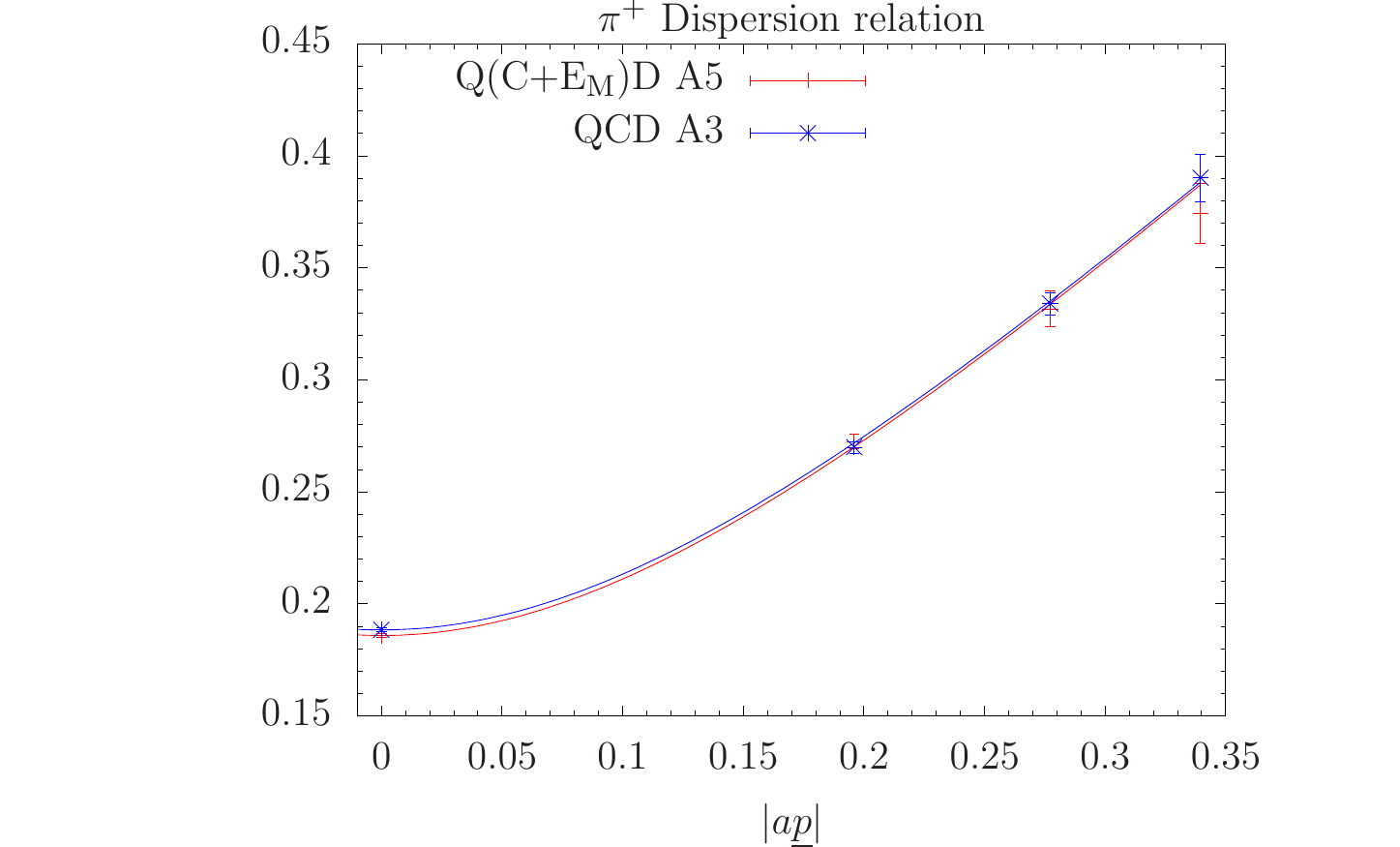}
\captionsetup{width=.95\linewidth}
\caption{Dispersion relations for
the lightest charged pseudoscalar meson in the two matched simulations.}
\label{fig:disp_rel_matched}
\end{subfigure}
\caption{Dispersion relation for pseudoscalar states with $m_\gamma = 0.1$. The solid lines represent
the expectation from the  continuum dispersion relation.}
\label{fig:disp_rel}
%\end{center}
\vspace{-0.7cm}
\end{figure}

In order to find a suitable range of photon masses we explored dif\mbox{}ferent values of $m_\gamma$.
It was emphasized in Ref.~\cite{Patella:2017fgk} that for \quotes{small} photon masses there could be a linear $t$-term in the ef\mbox{}fective energies for charged states. 
This term was studied and recognized in Ref.~\cite{Endres:2015gda} and explicitly subtracted by hand.
The presence of this term can be dangerous, because it may signal that the massive formulation is reducing to the so-called TL formulation\footnote{The TL formulation does not satisfy reflection positivity and therefore does not have a positive definite Hamiltonian, furthermore it is a non-local formulation.}.
In Fig.~\ref{fig:eff_masses_phot_mass} we show how dif\mbox{}ferent choices of photon masses af\mbox{}fect the charged pseudoscalar ef\mbox{}fective mass, for which we could not observe such a linearly rising term.
%\hspace{-1cm}
%%%%%%%%%%%%%%%%%%%%%%%%%%%%%%%%%%%%%%%%%%%%%%%%%%%%%%%%%%%%%%%%%%%%%%%%%
%%%%%%%%%%%%%%%%%%%%%%%%%%%%%%%%%%%%%%%%%%%%%%%%%%%%%%%%%%%%
\noindent \\
From Fig.~\ref{fig:eff_masses_phot_mass} we can draw a number of conclusions:
\begin{itemize}
\item QED$_{\rm L}$ is consistent with QED$_{\rm M}$ in the limit of $m_\gamma \rightarrow 0$. This suggests that the potential issues present in the L formulations are not af\mbox{}fecting spectroscopic quantities.
\item For large $m_\gamma$ the massive photon decouples and the results approach the QCD case.
\item As we discussed previously the inclusion of QED increases the quark masses and therefore the light pseudoscalar 
mesons get heavier (at fixed bare parameters).
\end{itemize}
Practically for photon masses $m_\gamma \lesssim 0.05$ the charged correlators become too small and it is impossible to reliable extract ef\mbox{}fective masses. That is a consequence of the reintroduction of the zero mode.
We checked that there is no dif\mbox{}ficulty in extracting the correct dispersion relation for masses $m_\gamma \gtrsim 0.05$, and that it agrees with the continuum one, see Fig.~\ref{fig:disp_rel}.
Furthermore the matching between the QCD and Q(C+E)D ensembles survives when we change the particle's momentum.
%%%%%%%%%%%%%%%%%%%%%%%%%%%%%%%%%%%%%%%%%%%%%%%%%%%%%%%%%%%%
%%%%%%%%%%%%%%%%%%%%%%%%%%%%%%%%%%%%%%%%%%%%%%%%%%%%%%%%%%%%
Finally we checked for final volume and photon mass ef\mbox{}fects and those are found to be negligible within errors.
\vspace{-0.2cm}
\section{Electromagnetic corrections to the muon anomaly}\label{em_corr_muon_anomaly}
The HVP tensor is given by 
\begin{align}
\Pi_{\mu\nu} (q) = \int\de^4 x \,\e^{iq\cdot x} \langle V_\mu(x) V_\nu(0)  \rangle,
\end{align}
where $V_\mu(x)$ is the quark electromagnetic current.
Such a  current is conserved in two-flavor Q(C+E)D as it is obtained from 
a combination of $\I$ and $\tau^3$ matrices in flavor space, i.e.
\begin{align}
V_\mu(x) = \Psibar(x) \gamma_\mu \left[    
\frac{Q_u}{2}\left(\I + \tau^3\right)
+ \frac{Q_d}{2}\left(\I - \tau^3\right)
\right]\Psi(x)
,
\end{align}
where $\Psi$ represents the flavor doublet $\left(u\,,d\right)^{\rm T}$.
The corresponding vector transformations are indeed preserved by electromagnetic interactions.
On the lattice we use the one-point-split current, that is exactly conserved, therefore the normalization constant $Z_V$ is one.
\subsection{Hadronic Vacuum Polarization}
For the calculation of the  HVP we neglect quark-disconnected diagrams, which are Zweig suppressed, 
and we recall that we are working in the electroquenched approximation.
Point sources are used throughout this preliminary study. 
%\\
We extract the scalar HVP from both the  diagonal components ($\mu=\nu$), after we have taken into account the contact term, 
and the non-diagonal ones ($\mu\neq\nu$).
We also employ the Zero Mode Subtraction (ZMS) modification of the tensor \cite{Bernecker:2011gh}.
Results are shown in Fig.~\ref{fig:hvp} where we plot the unsubtracted and subtracted scalar HVP as a function of $r_0^2 \hat{q}^2$, with $r_0$, from Ref.~\cite{Fritzsch:2012wq}, being the Sommer parameter and $\hat{q}_\mu = 2 \sin(q_\mu/2)$.
Notice that $r_0 /a$ as any other gluonic scale does not receive QED corrections in the quenched approximation.\\
In Fig.~\ref{fig:unsub_qecd_hvp} we see good agreement between the Q(C+E$_{\rm M}$)D and Q(C+E$_{\rm L}$)D results. 
Again, we interpret this as an indication that
 the L formulation, despite a number of theoretical issues, seems to provide a valid 
IR-regularization for the HVP.
\subsection{A new strategy}
\begin{figure}[!ht]
%\begin{center}
\begin{subfigure}{0.5\textwidth}
\includegraphics[scale=0.5]{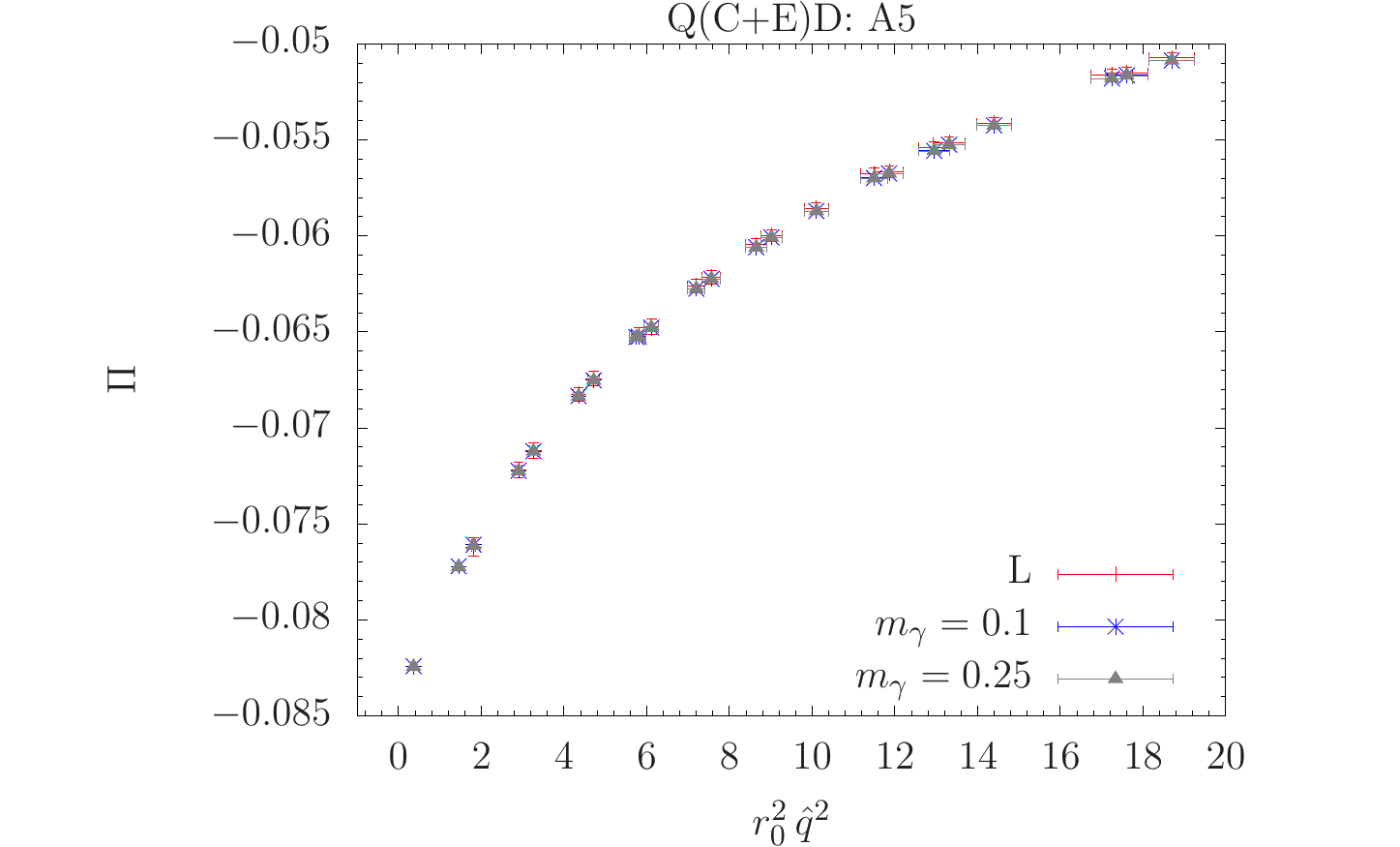} 
\captionsetup{width=.95\linewidth}
\caption{Unsubtracted HVP in Q(C+qE)D in A5 ensemble.}
\label{fig:unsub_qecd_hvp}
\end{subfigure}
%\hspace{-0.8cm}
\hfill
\begin{subfigure}{0.5\textwidth}
\includegraphics[scale=0.5]{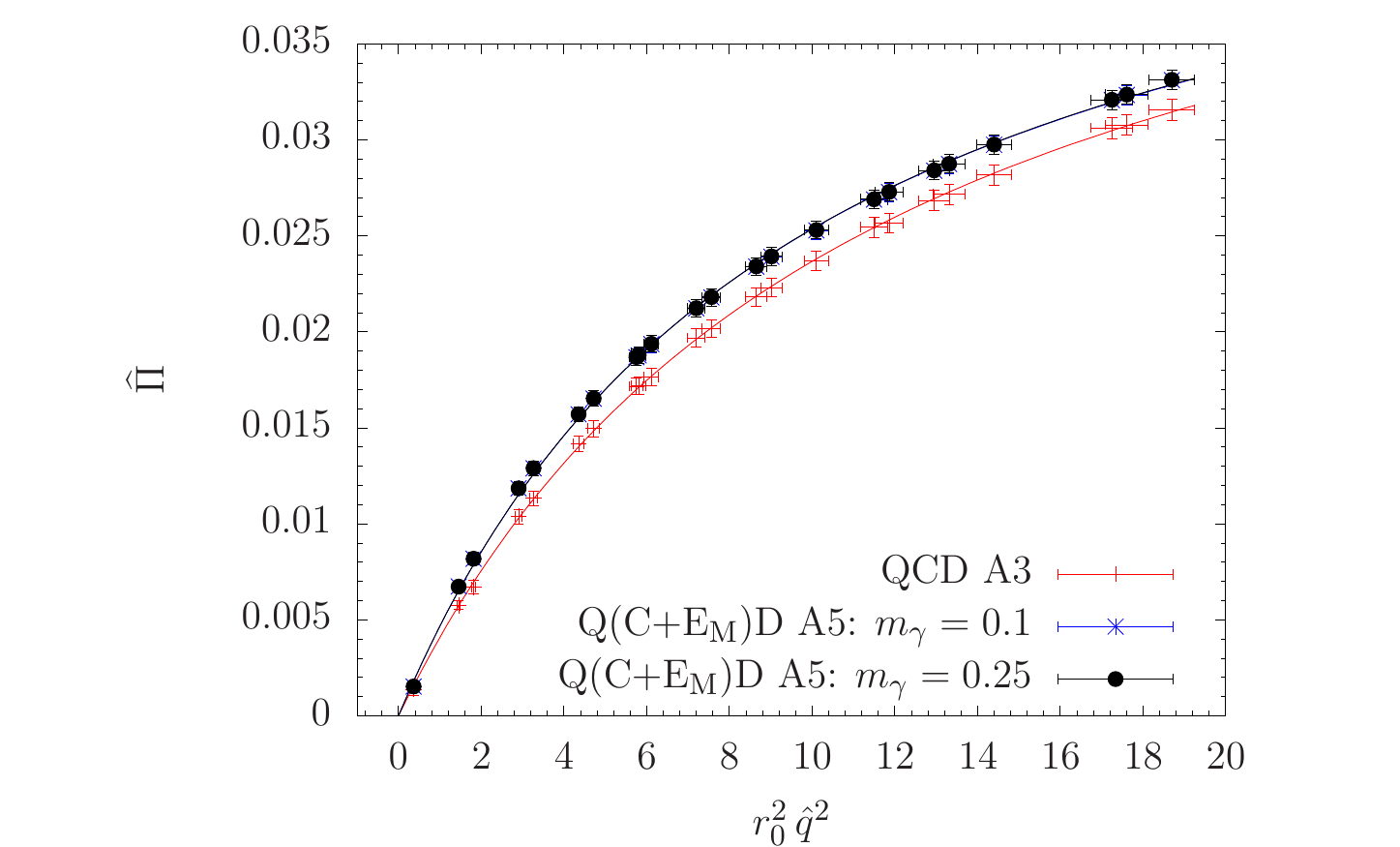}
\captionsetup{width=.95\linewidth}
\caption{Subtracted HVP with and without QED$_{\rm M}$.}
\label{fig:qecd_hvp_a5}
\end{subfigure}
\caption{HVP as a function of $r_0^2 \hat{q}^2$.}
\label{fig:hvp}
%\end{center}
\vspace{-0.7cm}
\end{figure}

\begin{figure}[!ht]
%\begin{center}
\begin{subfigure}{0.5\textwidth}
\includegraphics[scale=0.5]{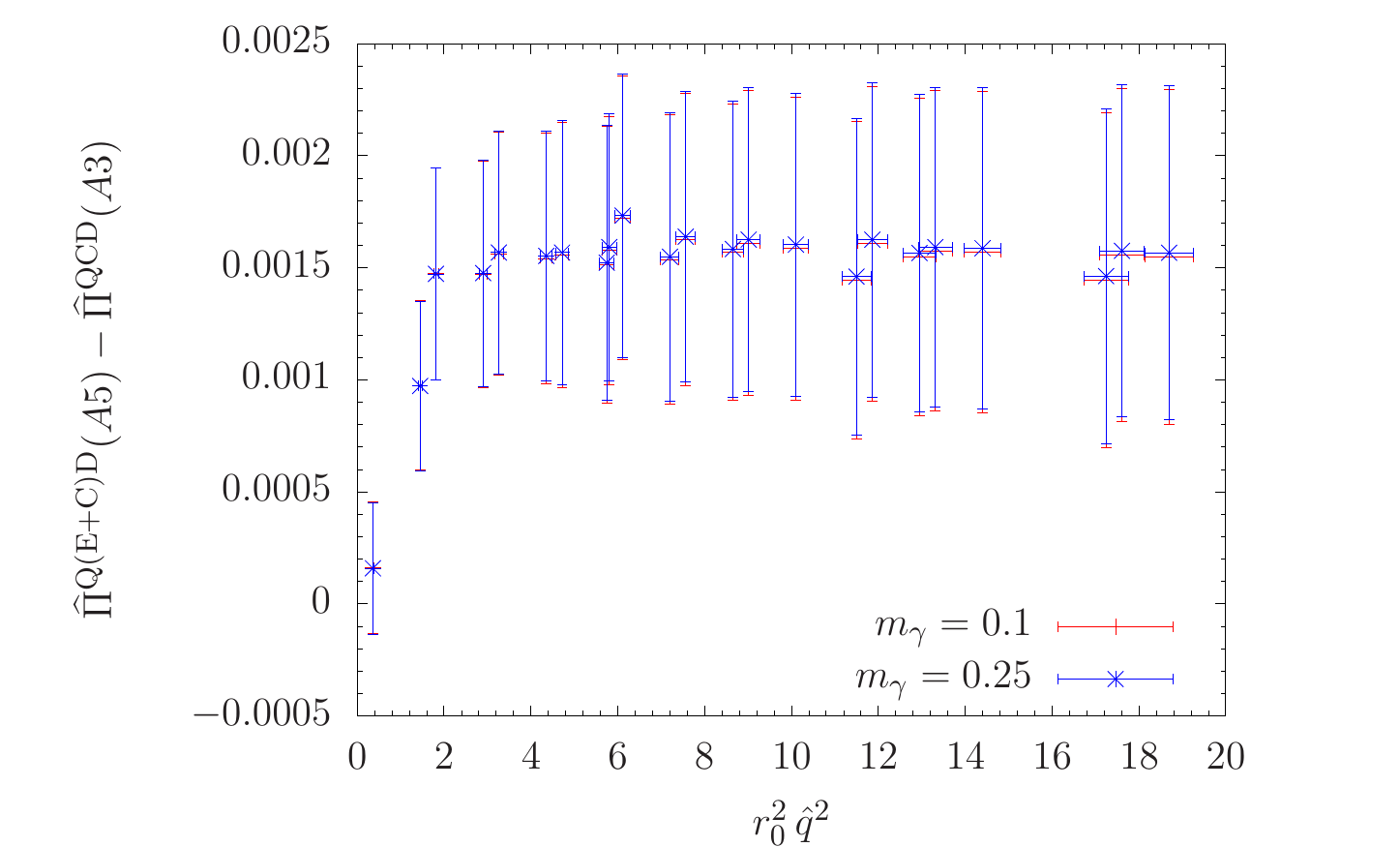} 
\captionsetup{width=.95\linewidth}
\caption{Dif\mbox{}ference of the HVP with and without QED$_{\rm M}$.}
\label{fig:conf_conf}
\end{subfigure}
%\hspace{-0.8cm}
\hfill
\begin{subfigure}{0.5\textwidth}
\includegraphics[scale=0.5]{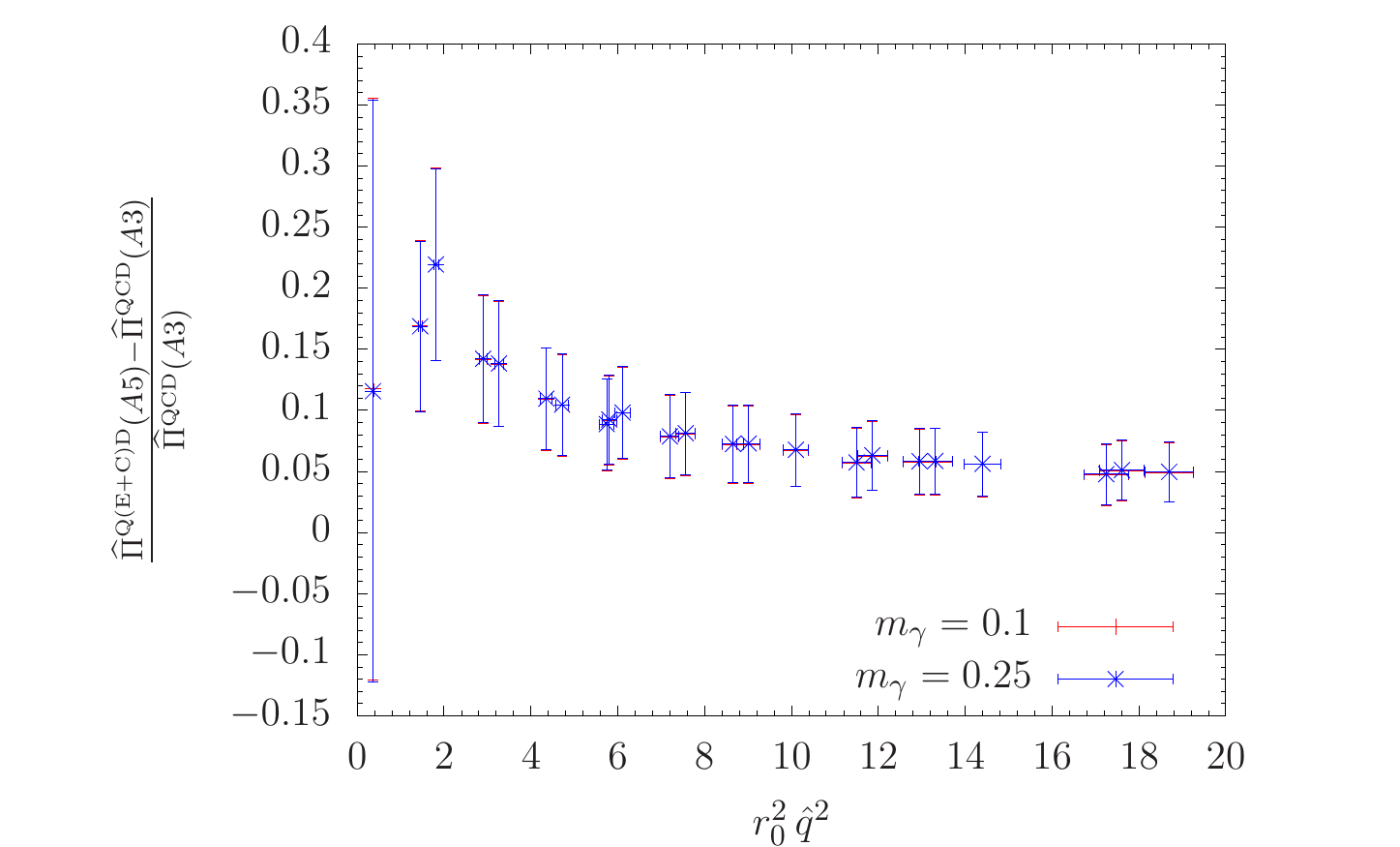}
\captionsetup{width=.95\linewidth}
\caption{Relative dif\mbox{}ference of the QCD and Q(C+E$_{\rm M}$)D HVP.}
\label{fig:conf}
\end{subfigure}
\caption{The EM contribution to the scalar VP $\delta \widehat{\Pi}$ as a function of $r_0^2 \hat{q}^2$.}
\label{fig:em_conf}
%\end{center}
\vspace{-0.7cm}
\end{figure}

Ideally we would like to compare the HVP with and without electromagnetic
ef\mbox{}fects.
The two HVPs will be dif\mbox{}ferent functions of the renormalized parameters, and to have
a meaningful comparison we need to consider them at the same renormalized parameters values.
One way to go could be to rescale the bare values for the change in the quark mass and the strong
coupling (reflected in the change of the lattice spacing), when
considering the electromagnetic corrections to the HVP. We expect all those changes to be at the
percent level. In this preliminary study we neglect the shift in the absolute scale.
In Ref.~\cite{DellaMorte:2017dyu} an estimate is provided, implying that a change of 1\% on the lattice spacing
is reflected in a $\simeq$ 1.5\% change in $a_\mu^{\rm HLO}=\left(\frac{\alpha}{\pi}\right)^2\int dq^2 f(q^2,m_\mu^2)\hat{\Pi}(q^2)$,
(see e.g., Ref.~\cite{Capitani:2015sba} for the definitions of $ f(q^2,m_\mu^2)$ and $\hat{\Pi}(q^2)$).
%For this reason we neglect the change in the lattice spacing and the change in the
%electromagnetic coupling. \\
A na\"ive strategy to compute the electromagnetic ef\mbox{}fects
on the muon anomaly would be:
\begin{itemize}
\item fit the scalar HVP in QCD and Q(C+E)D,
\item compute separately the muon anomaly $a_\mu$ in the two dif\mbox{}ferent theories,
\item take the resulting dif\mbox{}ference after the extrapolation to infinite volume, physical point and continuum.
\end{itemize} 
This procedure is quite inef\mbox{}ficient, since the EM ef\mbox{}fects can be easily washed out by the various systematics, e.g.~Pad\'e fit.\\
Our strategy makes use of the matching of the charged pion masses (traded for the quark masses) in the ensembles with QCD and Q(C+E)D.
%In this respect the change in the quark masses is already taken into account and the others are neglected.
In Fig.~\ref{fig:qecd_hvp_a5} we present the comparison between the QCD result and Q(C+E$_{\rm M}$)D one, that gives direct access to the electromagnetic ef\mbox{}fects.
%%%%%%%%%%%%%%%%%%%%%%%%%%%%%%%%%%%%%%%%%%%%%%%%%%%%%%%%%%%%
%%%%%%%%%%%%%%%%%%%%%%%%%%%%%%%%%%%%%%%%%%%%%%%%%%%%%%%%%%%
\noindent \\
Our strategy consists in:
\begin{itemize}
\item take the dif\mbox{}ference of the subtracted scalar HVPs, i.e.~$\delta \widehat{\Pi}\equiv \widehat{\Pi}^{\rm Q(C+E)D } - \widehat{\Pi}^{\rm QCD }$, at fixed pion mass,
\item fit $\delta \widehat{\Pi}$ and plug it in the formula for the anomaly $a_\mu^\delta \propto \int dq^2 f(q^2, m_\mu^2) \; \delta\hat{\Pi}(q^2)$.
Notice that the change in the lattice spacing between QCD and Q(C+E)D would be relevant here, and once that is considered the expression above is strictly
speaking no-longer valid, as one should consider two dif\mbox{}ferent $f$-kernels when converting the muon mass to lattice units.
\item finally extrapolate to infinite volume, physical point and continuum limit.
\end{itemize}
This consists of a one-fewer-fit procedure compared to the first strategy discussed and the systematics are 
reduced, as one can see in Fig.~\ref{fig:em_conf}. 
The signal is quite clear, even though compatible with zero within two $\sigma$s.
The slowly varying relative EM ef\mbox{}fect on the HVP could in fact be fitted by a constant around 7\%. 
%ef\mbox{}fect.
%%%%%%%%%%%%%%%%%%%%%%%%%%%%%%%%%%%%%%%%%%%%%%%%%%%%%%%%%%%%
%%%%%%%%%%%%%%%%%%%%%%%%%%%%%%%%%%%%%%%%%%%%%%%%%%%%%%%%%%%%
%\noindent \\
In view of that, in order to get an idea about the size of the EM corrections to $a_\mu^{\rm HLO}$, we decided not to perform
any fit and simply integrate numerically (using the trapezoidal rule) the function $f(q^2, m_\mu^2) \; \delta\hat{\Pi}(q^2)$ in $dq^2$
up to $r_0\hat{q}^2\simeq 20$. We obtain the result
%$\delta \widehat{\Pi}$ linearly up to $r_0\hat{q}^2_{\rm ref}$ and then with a constant from there on.
%We plug the resulting expression in the formula in the second bullet-point above and obtain the result
\begin{align}
a_\mu^\delta \times 10^{10}= 21\pm 9_{\rm stat} \,,
%\begin{cases}
%68 \pm 28 \quad\text{ for } R_{10}\\
%50 \pm 24 \quad\text{ for } R_{11}\\
%35 \pm 14 \quad\text{ for lin.-const. } 
%\end{cases},
\end{align}
where the error is statistical only.
%
%obtained as the difference between the results for  $r_0\hat{q}^2_{\rm ref}=1.5$ and  $r_0\hat{q}^2_{\rm ref}=2.5$.  
%where $R_{10}$ and $R_{11}$ represents the Pad\'e fits and lin.-const.~represent a linear fit in the low $r_0^2 \hat{q}^2$ region and a constant afterwards.
The ef\mbox{}fect seems to be of the same size  as  the theo-exp discrepancy.
There are still important sytematic ef\mbox{}fects to be quantified though, as the change in the lattice spacing and in the pion masses.
\vspace{-0.3cm}
\section{Conclusions}
We presented a new way to isolate electromagnetic ef\mbox{}fects for the
hadronic contribution to $(g-2)_\mu$. We added quenched QED configurations to preexistent QCD
ones. We considered two dif\mbox{}ferent regularizations of the finite volume zero modes.
 We were able
to see a clear ef\mbox{}fect, even for physical quark charges and electromagnetic coupling. 
The crucial
question to be answered in the future is whether the ef\mbox{}fect we have seen is going to be larger for
smaller pion masses or not. 
Within the limitations of the computation (single lattice spacing, pion
mass around 400 MeV), we saw an ef\mbox{}fect of the same size as the discrepancy in $a_\mu$ between
theory and experiments.\\
In order to reduce the systematics, we are planning to analyze dif\mbox{}ferent
volumes, pion masses and lattice spacings, already available within the CLS initiative. 
Other
improvements can be achieved by considering the addition of a clover term for the electromagnetic
part of the action, as well as improving the vector current, in order to have better control on the
continuum extrapolation. We may also have to consider reweighting in the bare gauge coupling in order
to match lattice spacings (between QCD and Q(C+E)D) and isolate electromagnetic corrections 
following the approach we described.

The main extension in order to properly assess isospin breaking corrections remains however
the inclusion of the up-down quarks mass splitting (strong isospin breaking). Work in that
direction is progressing.

%Another crucial ef\mbox{}fect to be included in the hadronic muon anomaly in the
%future is the quark mass splitting.
% This ef\mbox{}fect can go in the opposite direction of the EM ones,
%an example of that is visible in the proton-neutron mass splitting.

\noindent
{\bf Acknowledgements.}
We wish to thank Agostino Patella for useful discussions and Georg von Hippel for help
in accessing the CLS configurations.
This work was supported by the Danish National Research Foundation DNRF:90 grant and by
a Lundbeck Foundation Fellowship grant. The computing facilities were provided by the Danish
Centre for Scientific Computing and the DeIC national HPC center at SDU.

\bibliography{lattice2017}

\end{document}